\documentclass[twocolumn,showpacs,preprintnumbers,amsmath,amssymb]{revtex4}

\begin{document}

\title{$T$-odd quark distributions: QCD versus chiral models}
\author{P.V.~Pobylitsa}
\affiliation{
Institute for Theoretical Physics II, Ruhr University Bochum,
D-44780 Bochum, Germany\\
and\\
Petersburg Nuclear Physics Institute, Gatchina, St. Petersburg, 188300,
Russia}

\begin{abstract}
The $T$-odd quark distrbutions are shown to vanish in the chiral sigma model in
contrast to the opposite widespread opinion. This failure of the chiral
sigma model is a feature of the model itself and has nothing to do with the
recent progress in the clarification of the status of the $T$-odd
distributions in QCD.
\end{abstract}

\maketitle

Recently certain progress has been made \cite{BHS-02,Collins-02,JY-02,BJY-02}
in understanding the properties of the so-called $T$-odd quark distributions
in QCD \cite{Sivers-90,ABM-95,BM-98,BM-99}. Naively the $T$-odd quark
distribution functions in nucleon are defined as follows (notation of Ref.~\cite{ABDM-02}): 
\[
\Delta_{0}^{T}f(x,{\bf k}_{\perp}^{2})=\mathrm{Im}\int\frac{dy^{-}d^{2}{\bf y}_{\perp}}{
2(2\pi)^{3}}e^{-ixP^{+}y^{-}+i\mathbf{k}_{\perp}\cdot\mathbf{y}_{\perp}}
\]
\begin{equation}
\times\langle P,-|\bar{\psi}(0,y^{-},{\bf y}_{\perp})\gamma^{+}\psi(0)|P,+\rangle
\,,  \label{def-naive-1}
\end{equation}
\[
\Delta_{T}^{0}f(x,{\bf k}_{\perp}^{2})=\int\frac{dy^{-}d^{2}{\bf y}_{\perp}}{2(2\pi)^{3}}
e^{-ixP^{+}y^{-}+i\mathbf{k}_{\perp}\cdot\mathbf{y}_{\perp}}
\]
\begin{equation}
\times\langle P|\bar{\psi}(0,y^{-},{\bf y}_{\perp})i\sigma^{2+}\gamma_{5}\psi(0)|P
\rangle.  \label{def-naive-2}
\end{equation}
Here $\psi$ is the quark field and $|P,\pm\rangle\,$is the nucleon state
with the given momentum and helicity.

If one uses this \emph{naive definition} of functions $\Delta_{0}^{T}f$
(introduced by Sivers \cite{Sivers-90}) and
$\Delta_{T}^{0}f$ (studied in Ref. \cite{BM-98}),
then \emph{in any theory} with nonbroken $T$-invariance
these functions should vanish. However, as noticed in Ref.~\cite{Collins-02}
the \emph{correct QCD definition} of functions $\Delta_{0}^{T}f$, $\Delta
_{T}^{0}f$ should include properly chosen Wilson lines. Moreover, the
direction of the Wilson lines is sensitive to the considered hard process. The
appearance of the Wilson lines in the definition of functions $\Delta_{0}^{T}f$, 
$\Delta_{T}^{0}f$ breaks the naive $T$-invariance argument so that one can
have nonvanishing functions $\Delta_{0}^{T}f$, $\Delta_{T}^{0}f$ in QCD.

In Ref.~\cite{ABDM-02} an attempt was made to study $T$-odd distributions in
the chiral model with the lagrangian
\[
L=i\bar{\psi}\gamma^{\mu}\partial_{\mu}\psi-g\bar{\psi}\left( \sigma
+i\gamma_{5}\tau^{a}\pi^{a}\right) \psi
\]
\begin{equation}
+\frac{1}{2}(\partial_{\mu}\sigma)^{2}+\frac{1}{2}(\partial_{\mu}
\pi^{a})^{2}-U(\sigma,\pi)  \label{L-sigma}
\end{equation}
with the mexican-hat potential $U$. Actually most of the arguments of
Ref.~\cite{ABDM-02} and of this paper are valid for a wider class of models with
the action
\[
S[\psi,\sigma,\pi]=\sum\limits_{c=1}^{N_{c}}\int d^{4}x
\]
\begin{equation}
\times\left[ i\bar{\psi}
_{c}\gamma^{\mu}\partial_{\mu}\psi_{c}-g\bar{\psi}_{c}\left( \sigma
+i\gamma_{5}\tau^{a}\pi^{a}\right) \psi_{c}\right] 
+N_{c}S_{1}(\sigma ,\pi)
\label{CQSM}
\end{equation}
containing a chiral invariant mesonic piece $S_{1}(\sigma,\pi)$. We have
explicitly written the summation over the quark color indices $c$ and have
inserted the number of colors $N_{c}$ in front of the bosonic part of the
action in the form which allows a systematic $1/N_{c}$ expansion in the
large-$N_{c}$ limit. It is assumed that the parameters of the chiral model
are chosen so that the chiral $SU(2)_{L}\otimes SU(2)_{R}\otimes U(1)_{V}$
symmetry is spontaneously broken to the vector symmetry group $U(2)_{V}$ and
that the model contains states with the nucleon quantum numbers. We also
assume that the action $S_{1}(\sigma,\pi)$ is $C$, $P$ and $T$ invariant.

The main conclusion of Ref.~\cite{ABDM-02} is that the $T$-odd
quark distribution functions computed in the model (\ref{L-sigma}) might be
nonvanishing. This result is rather strange. Indeed, the lagrangian
(\ref{L-sigma}) is explicitly $T$-invariant, so that the theory is invariant
under the \emph{standard} time reflection (STR) transformations of the
quantum fields:
\begin{eqnarray}
T\psi(x^{0},\mathbf{x})T^{-1} & =&C\gamma_{5}\psi(-x^{0},\mathbf{x})\,, 
\nonumber \\
T\sigma(x^{0},\mathbf{x})T^{-1} & =&\sigma(-x^{0},\mathbf{x})\,,  \label{STR}
\\
T\pi^{a}(x^{0},\mathbf{x})T^{-1} & =&\varepsilon^{a}\pi^{a}(-x^{0},\mathbf{x})\,,
\nonumber
\end{eqnarray}
\begin{equation}
\varepsilon^{1}=\varepsilon^{3}=-1,\,\varepsilon ^{2}=1\,.
\end{equation}

Under the following conditions:

\begin{itemize}
\item the chiral model makes sense (the ground state is stable and the
states with the quantum numbers of nucleon exist),

\item no exotic phenomena happen like spontaneous breakdown of the $T$-invariance,

\item one uses the naive definition of the $T$-odd functions
(\ref{def-naive-1}), (\ref{def-naive-2}),
\end{itemize}
the STR symmetry $T$ (\ref{STR}) of the model (\ref{L-sigma})
automatically leads to the conclusion that the $T$-odd functions
(\ref{def-naive-1}), (\ref{def-naive-2}) vanish in this model in contrast to the
argument of Ref.~\cite{ABDM-02}.

Let us try to understand what stands behind this discrepancy. First one
notices that the authors of Ref.~\cite{ABDM-02} actually do not compute the
$T$-odd parton distributions. Their conclusion about nonvanishing $T$-odd
distributions is based only on certain symmetry arguments.

Instead of the STR transformation $T$ (\ref{STR}) the authors of
Ref.~\cite{ABDM-02} suggest to consider what they call
``nonstandard time reversal''
(NSTR) transformation $T_{\mathrm{NSTR}}$ which differs from STR $T$ simply by the isotopic
180$^\circ$ rotation: 
\begin{equation}
T_{\mathrm{NSTR}}=e^{i\pi I_{2}}T  \label{NSTR}
\end{equation}
so that 
\begin{eqnarray}
T_{\mathrm{NSTR}}\psi(x^{0},\mathbf{x})T_{\mathrm{NSTR}}^{-1} & =&-i\tau
_{2}C\gamma_{5}\psi(-x^{0},\mathbf{x})\,,  \nonumber  \\
T_{\mathrm{NSTR}}\sigma(x^{0},\mathbf{x})T_{\mathrm{NSTR}}^{-1} &
=&\sigma(-x^{0},\mathbf{x})\,, \\
T_{\mathrm{NSTR}}\pi^{a}(x^{0},\mathbf{x})T_{\mathrm{NSTR}}^{-1} & =&\pi
^{a}(-x^{0},\mathbf{x})\,.  \nonumber 
\end{eqnarray}

Since the isotopic invariance is a symmetry of the model
(\ref{L-sigma}), it does
not matter whether one works with STR or NSTR. But one should not forget
that NSTR is not the only symmetry of the theory, there is also the isotopic
symmetry. However, the authors of Ref.~\cite{ABDM-02} concentrate on NSTR and
ignore other symmetries. They make the correct observation that NSTR taken
alone allows nonzero $T$-odd functions and they declare without any
calculation that the sigma model may lead to nonzero $T$-odd distributions. The
fault of this argument is that NSTR is not the only symmetry of the model
--- if one combines NSTR with the isotopic invariance to recover STR, then
one can use the standard argument to show that the $T$-odd functions vanish.

One could wonder, why in Ref.~\cite{ABDM-02} such a significance is
attributed to NSTR, whereas the isotopic invariance and STR are ignored,
although they are exact symmetries of the model. It seems that the authors
mix two different issues:

1) the symmetries of the action which are not spontaneously broken (these
symmetries include both STR and NSTR as well as the isotopic invariance),

2) the symmetries of the mean field describing the nucleon states.

As it is well known, the chiral quark-soliton models of type (\ref{CQSM})
allow a systematic $1/N_{c}$ expansion in the limit of the large number of
colors $N_{c}$ \cite{Witten-79,GS-84}. The leading order of this $1/N_{c}$
expansion is nothing else but the mean field approximation. In terms of the
path integral approach, the $1/N_{c}$ expansion can be constructed as a
systematic expansion about the corresponding saddle point of the effective
action, which can be obtained by the Gaussian integration over the quark
fields. The properties of the vacuum, in particular the spontaneous
breakdown of the chiral symmetry, are controlled by the constant (space and
time independent) saddle point $\sigma_{0},\pi_{0}^{a}$. As for the nucleon
states, they are described by the $\mathbf{x}$ dependent (but static) saddle
point usually referred to as ``soliton''. This soliton field has the form 
\begin{eqnarray}
\sigma_{\mathrm{sol}}(\mathbf{x}) & =f_{1}(|\mathbf{x}|)\,,  \nonumber  \\
\pi_{\mathrm{sol}}^{a}(\mathbf{x}) & =x^{a}f_{2}(|\mathbf{x}|)\,,
\end{eqnarray}
which violates a lot of symmetries: spatial translations and $O(3)$
rotations, $SU(2)$ isotopic rotations, STR. On the other hand, the soliton
field is invariant with respect to combined space-isotopic rotations and
with respect to NSTR. Probably due to this privileged status of NSTR, the
authors of Ref.~\cite{ABDM-02} have concentrated on this symmetry ignoring
the others. But the symmetries of the soliton field have nothing to do with
the symmetries of the Hamiltonian and of the ground state. For example,
the soliton field is not translationally invariant but this does not mean that
the translational invariance of the model is broken.

Thus both STR and NSTR are exact symmetries of the model. Another way to
understand that STR is an exact symmetry is to remember that according to
Eq.~(\ref{NSTR}) NSTR differs from STR by an isotopic rotation. If STR were
broken (spontaneously or in some other way) with NSTR remaining unbroken,
then this would automatically lead to the broken isotopic symmetry.
The third argument in favor of STR comes from the \emph{CPT}
theorem: the breakdown of STR would mean that the $CP$
symmetry is violated in the model
(\ref{L-sigma}), which is probably not the aim of the authors of
Ref.~\cite{ABDM-02}.

The illusion of the violation of STR by the soliton field actually
disappears if one properly uses the standard methods of the treatment of
collective coordinates in the semiclassical quantization \cite{GS-84,GS-7x}.
Using this technique, one can in principle check the STR symmetry
explicitly in any order of the $1/N_{c}$ expansion.

To summarize, as long as the isotopic $SU(2)$ invariance is not broken,
there is no difference whether one uses STR or NSTR --- anyway
the $T$-odd distribution functions (\ref{def-naive-1}), (\ref{def-naive-2})
vanish in the model considered in Ref.~\cite{ABDM-02} in contrast to the
conclusion of the authors.

From the above analysis it is obvious that the vanishing of $T$-odd
distributions (\ref{def-naive-1}), (\ref{def-naive-2}) in the model
(\ref{L-sigma}) is an internal feature of this model and has nothing to do with
the nonvanishing $T$-odd distributions in QCD.

The reason allowing the existence of the $T$-odd parton distributions in QCD
is the nontrivial role played by the direction of the Wilson lines accompanying the quark
fields~\cite{Collins-02}.
If one really wants to model the QCD effects, then one has to find a
way to incorporate the gauge links into the effective model.

On general grounds it is expected that the large-$N_{c}$ QCD in the color
singlet sector should be equivalent to some effective theory with (generally
bilocal) meson fields \cite{Witten-79}. However, the corresponding action is
not known even in the leading order of the $1/N_{c}$ expansion. Instead one
usually deals with ad hoc models like the sigma model (\ref{L-sigma}). In
the best case these models can reproduce some features of QCD like
spontaneous breakdown of chiral symmetry, correct large-$N_{c}$ counting,
certain pieces of the hadronic spectrum etc. Since these models are not
derived from QCD, the question how to implement the Wilson lines in these
models can hardly be treated seriously.

An interesting attempt to describe the gluonic degrees of freedom within a
low energy effective model was made in the framework of the instanton
vacuum model \cite{DP-86}. Although the ``derivation'' of this model from
QCD involves quite a number of approximations and simplifications, in
principle one can trace the relation between the gluonic degrees of freedom
of QCD and the quark-pion fields of the effective theory \cite{DPW}. It
would be interesting whether within this approach one could say something
about the $T $-odd quark distribution functions.

\textbf{Acknowledgement.} The author appreciates discussions with
Ya.~I.~Azimov, D.~Boer, A.~Metz, P.~J.~Mulders, M.~V.~Polyakov
and P.~Schweitzer.

\end{document}